\documentclass[authoryear,fleqn,5p]{elsarticle} 

\journal{Energy Policy}

\usepackage{tikz}
  \usepackage{pgfplots,pgfplotstable}
  \usetikzlibrary{pgfplots.groupplots}
  \pgfplotsset{compat=1.5}
  
\usepackage{cuted} 

\usepackage{booktabs}
\usepackage{multirow}

\usepackage{array}
\newcolumntype{M}[1]{>{\centering\arraybackslash}m{#1}}

\usepackage{amsmath}
\usepackage{amssymb}
\usepackage{mathtools}
\usepackage{cuted}
\usepackage{graphicx}
\usepackage{flafter}
\usepackage{placeins}
\usepackage{textcomp}
\usepackage{hyperref}
\usepackage{xurl} 
\usepackage{color}
\hypersetup{urlcolor=cyan}

\usepackage{flushend} 

\usepackage{subcaption} 

\usepackage{enumitem} 

\usepackage{ragged2e} 

\makeatletter
\def\ps@pprintTitle{%
  \let\@oddhead\@empty
  \let\@evenhead\@empty
  \def\@oddfoot{\reset@font\hfil\thepage\hfil}
  \let\@evenfoot\@oddfoot
}
\makeatother

\begin{document}

\setlength{\tabcolsep}{6pt}
\setlength{\mathindent}{0pt}

\begin{frontmatter}

\title{Who should pay for frequency-containment ancillary services? \\Making responsible units bear the cost to shape investment in generation and loads}

\author[UPM]{Luis Badesa\corref{cor1}}
\ead{luis.badesa@upm.es}
\author[Imperial]{Carlos Matamala}
\author[Imperial]{Goran Strbac}
\address[UPM]{Technical University of Madrid (UPM), Ronda de Valencia 3, 28012 Madrid, Spain}
\address[Imperial]{Imperial College London, South Kensington, London, SW7 2AZ, UK
\vspace*{-4mm}}

\cortext[cor1]{Corresponding author}

\begin{abstract}
While the operating cost of electricity grids based on thermal generation was largely driven by the cost of fuel, as renewable penetration increases, ancillary services represent an increasingly large proportion of the running costs. Electric frequency is an important magnitude in highly renewable grids, as it becomes more volatile and therefore the cost related to maintaining it within safe bounds has significantly increased. So far, costs for frequency-containment ancillary services have been socialised in most countries, but it has become relevant to rethink this regulatory arrangement. In this paper, we discuss the issue of cost allocation for these services, highlighting the need to evolve towards a causation-based regulatory framework. We argue that parties responsible for creating the need for ancillary services should bear these costs. However, this would imply an important change in electricity market policy, therefore it is necessary to understand the impact on current and future investments on generation, as well as on electricity tariffs. Here we provide a mostly qualitative analysis of this issue, defining guidelines for practical implementation and further study. 
\end{abstract}


\end{frontmatter}

\section{Introduction} \label{sec:Intro}

This paper addresses the issue of increasing costs for frequency-containment services in electricity grids with a high renewable penetration, and how to appropriately distribute the cost among the agents that create the need for these services. The goal of rethinking the distribution of expenses is to achieve a reduction in the overall cost incurred in frequency security. Eventually, the savings would be reflected on electricity tariffs, 
decreasing the price paid by consumers for grid-management services.

Currently, frequency-containment costs are socialised in most countries, i.e., a flat tariff is applied to all consumers, as is the case in Great Britain \citep{BSUoS}. However, we argue here that cost allocation should follow cost causation: 
those who create the problem should bear the costs arising from fixing it. It is not a matter of fairness, but rather an instrument to encourage the responsible parties in the power system to do as little `harm' as possible to the overall grid. By internalising these costs, the actors that cause the need for ancillary services would search for more cost-effective options to reduce this need, therefore benefitting the electricity sector as a whole. An explanation on who these relevant actors are is provided in Section~\ref{sec:Challenge}.

Extensive previous work has been done on how to allocate the cost of different assets and services in power systems among users of the grid. Techniques from cooperative game theory have been applied for this purpose.
For the transmission network, \cite{zolezzi2002transmission} studied how much different grid users should pay for existing transmission lines, using the Shapley value and nucleolus methodologies.
In \cite{kristiansen2018mechanism}, a fair allocation of expected net benefits associated with new lines was proposed, also using the Shapley value.

Regarding cost allocation for frequency ancillary services, most work has focused on frequency regulation, i.e, constantly regulating electric frequency due to minor random imbalances in the grid, typically caused by demand fluctuations. A customer-specific distribution of costs using a proportional rule was developed by \cite{KirbyRegulation} for this service.
In \cite{ChuChen_AllocationPFR}, the cost of frequency reserves is distributed among loads according to their size and variability. The proposed method takes into account the correlation between demand fluctuations and the need for frequency reserves, where loads with higher variability pay a higher share of the costs.
Recent work by \cite{Liu_AllocationMilleage} introduced a cost allocation method for regulation mileage costs, using a linear rule to allocate cost to each market participant in proportion to their contribution to the system imbalance. 
However, none of these methods consider the desirable properties for an effective cost allocation mechanism, such as the avoidance of cross-subsidies among market players.

Proposals for cost allocation in power systems that do provide guarantees for desirable economic properties include \cite{HuChen_AllocationST}, where Shapley value and nucleolus strategies were used to distribute the costs associated with non-convexities, such as the start-up cost of thermal generators. These guarantee an equitable distribution of costs among consumers, improving the existing socialisation based on applying proportional rules. A sequential approach, equivalent to the Shapley value, was used in \cite{hirst2003allocating} to allocate contingency reserves costs among generators. However, this method cannot be directly applied to frequency-containment ancillary services: 
unlike traditional contingency reserves, the cost of frequency-containment services currently exhibits a nonlinear relationship with the size of the contingency. This is due to the effect of decreasing inertia in low-carbon power grids, which implies that the rule for determining the volume of frequency reserves necessary must be deduced from a differential equation, not a linear steady-state relationship as for contingency reserves. 

The study we present here builds upon the recent work in \citet{CarlosAllocation}. In the present paper, we analyse the practical implications that a new regulatory framework for frequency containment could have on existing and future power systems. We address several important questions, in order to shed some light on the main features that the cost allocation mechanism should have, and identify the primary aspects that should be considered before enforcing the shift in policy. For example, it is relevant to define if this cost allocation should apply to existing market players, or only to new entrants. Alternative arrangements could also be proposed, such as forcing renewable generators to provide an inertia-like support via a grid code, with the aim of reducing overall frequency-containment costs. These points are addressed in Section~\ref{sec:Allocation}, while a numerical example is included in Section~\ref{sec:TestCase} to illustrate the functioning of our proposed cost allocation mechanism.

Furthermore, in Section~\ref{sec:Impact} we focus on potential impacts that this new market structure would have on the business case of future zero-emissions generation, notably nuclear and offshore wind. Mitigation options for reducing their allocated cost are identified, as well as for large industrial consumers, who could also be responsible for a significant proportion of ancillary services costs. Finally, we identify outstanding questions to be resolved before this new framework may become a reality. 

With this study, we intend to provide the starting point for a debate on the specific scheme that should be put in place. Given the urgency of decarbonising, and the potential benefits that this cost-reflective framework could have on shaping investment during the energy transition as well as in future fully-decarbonised grids, the debate must happen now.

\subsection{The challenge of frequency security in low- and zero-carbon electricity grids} \label{sec:Challenge}

Before starting the discussion on the allocation of costs, let us give a brief summary on the main variables that influence frequency in an electricity grid.

\begin{figure}[!t]
	\centering
    \begin{subfigure}[b]{0.5\textwidth}
        \hspace*{-8mm}
        \centering
        \includegraphics[width=3.5in]{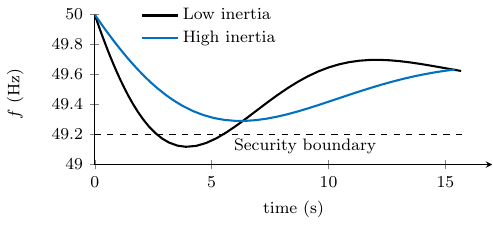}
        \captionsetup{width=0.95\linewidth} 
        \caption{\footnotesize Cases of low (100 GWs) and high inertia (200 GWs), both for a contingency of 1.8 GW.}
        \vspace{3mm}
        \label{fig:FreqInertia}
    \end{subfigure}%
    ~ 
    \newline 
    \begin{subfigure}[b]{0.5\textwidth}
        \hspace*{-8mm}
        \centering
        \includegraphics[width=3.5in]{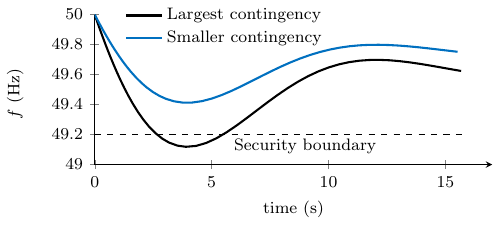}
        \captionsetup{width=.95\linewidth} 
        \caption{\footnotesize Cases of the largest (1.8 GW) and a smaller contingency (1.2 GW), both for a level of inertia of 100 GWs.}
        \label{fig:FreqContingency}
    \end{subfigure}
    \caption{Frequency excursions following the contingency (i.e., the sudden disconnection) of a generator in the power grid. The impact of inertia and contingency size is shown, for several cases representative of the British system.}
    \label{fig:FreqExcursions}
\end{figure}

The electric frequency of the grid reflects the balance between generation and demand. If the two magnitudes are perfectly balanced, frequency stays constant at the nominal value of 50 Hz (60 Hz in some grids, such as North America). However, if demand exceeds generation, frequency starts to decrease. This happens because synchronous generators, such as coal- and gas-fired power plants, start to slow down, as they release the kinetic energy stored in their rotating masses to compensate for the lack of power on the generation side. It is precisely the rotating speed of synchronous generators which sets the electric frequency of the grid. Note that the opposite is true for situations when generation exceeds demand, i.e., frequency would rise since synchronous generators would speed up to store more kinetic energy, therefore momentarily increasing demand in the grid.

The cost of procuring frequency-containment services has increased in recent years due to the rise in renewable generation, as highlighted by the low-carbon period during the COVID-19 lockdown in Great Britain, when these costs increased three times compared to normal conditions \citep{LuisCovid}. The higher cost is due to the decrease in system inertia, driven by the replacement of the rotating masses in thermal generators by the power-electronics-controlled wind turbines and solar photovoltaics. Given that inertia naturally acts as a buffer of kinetic energy, which helps stabilise the grid after a sudden generation-demand imbalance, alternative services must be procured. This implies that higher volumes of frequency reserves are necessary, that is, fast power injections from part-loaded generators or storage. These reserves are typically procured by transmission system operators via auctions, or by forcing generators to provide these services via grid codes.

The effect of a lower inertia on a large frequency excursion is shown in Figure~\ref{fig:FreqInertia}, which displays the evolution of grid frequency `$f$' after a loss of generation. The graph was obtained from a simplified model representative of the British grid, as described in \citet{LuisEFR}. The simulations were run with all other parameters being equal, such as the frequency reserves controls, demonstrating that a lower inertia implies a faster and larger frequency decline. In this instance, frequency in the low-inertia case drops below the security boundary in Great Britain, which is of 49.2 Hz. Some load would be disconnected, as frequency-sensitive relays deployed across the grid are adjusted to trip at this frequency threshold, in order to restore the power balance. 

Another key magnitude for frequency security is the size of the contingency. Figure~\ref{fig:FreqContingency} shows that the effect of a large contingency may be even more relevant to that of a lower inertia: low inertia makes frequency more volatile following a contingency, but if this contingency is small, the frequency excursion will not be very significant even in a low-inertia system. Given that the simultaneous outage of two different generators or loads is considered to have a negligible probability of occurrence, system operators always determine the volume of frequency reserves necessary based on the largest generator and load: if there are sufficient services to cover the largest single-unit contingency, any smaller event will also be secured. 

Therefore, we can conclude that a system composed by a few large units is weaker than another system of the same size formed by a large number of small units, since losing any single unit in the latter system would only cause a negligible imbalance. Large units, either generators or loads, are the actors creating the need for frequency-containment services in the first place. On the other hand, a power grid where the largest possible imbalance is small could operate safely with low inertia, even if procuring low volumes of frequency reserves.

\subsection{International experience on cost allocation of frequency-containment services} 

Frequency-containment services are most relevant for power grids in islands, either geographical or electrical. These grids typically have no alternating-current connections to a larger, continental grid, and therefore they are already suffering low inertia conditions driven by decarbonisation efforts. Three notable cases are Great Britain, Texas and Australia. However, while markets for these services have seen significant development in all three systems in recent years, aiming to achieve a more efficient procurement, neither Great Britain nor Texas apply a distribution of costs among responsible actors.

Australia is currently the only country applying a cost allocation based on cost causation to frequency-containment ancillary services. It is applied to both frequency reserves to raise power (to compensate for a generator contingency), and reserves to lower power (compensating load disconnection), the cost of the former being covered by generators and the latter by consumers \citep{PRAKASH2022112303}.

Two different methods are in fact used to compute the distribution of costs. This is a consequence of the Australian Energy Market Operator being responsible for managing two independent grids: the main Australian grid, called the `National Electricity Market' (NEM) and operating in the East coast, 
and a significantly smaller system called the `Wholesale Electricity Market' (WEM), in the Southwestern coast, with a peak demand of just over 4~GW.
A proportional cost allocation rule is used in the NEM \citep{AEMO2023_costAllo}, having the important problem of maintaining cross-subsidies, as mentioned at the beginning of Section~\ref{sec:Intro} and further considered in Section~\ref{sec:TestCase}. On the other hand, the WEM applies a sequential cost allocation \citep{AEMO2023_MarkerDesign_WEM}. As discussed later in the present work, this is a suitable cost-sharing rule, making responsible units appropriately cover the share of frequency-containment costs that they create.

The contribution of this article is to provide a theoretical analysis of the need for and desirable structure of a cost allocation framework, which is largely in agreement with the current practice in Australia's WEM. This study, which was not available in the literature, provides the basis for exporting this mechanism to other countries.

\section{Distributing costs of ancillary services among market players} \label{sec:Allocation}

In this section, we discuss the reasoning behind the cost-allocation mechanism as well as its practical implementation. We do so by explaining the procedure to be used for splitting the ancillary services cost among market players, and by addressing several questions on alternative cost-sharing designs.

\subsection{Why not keep socialising costs?}

Ancillary services costs are currently socialised in most countries, an acceptable arrangement until recently given that they represented a minor proportion of overall system costs. The limited benefits that a causation-based cost allocation would bring did not justify overcoming the challenges of enforcing the shift in regulation. However, the rise of renewables has changed this situation, as most electricity production will no longer require the use of fuel (therefore reducing running costs in the power system), while new stability challenges appear in the grid. This makes ancillary services increasingly relevant, and their associated costs, higher.

\vspace{1mm}

\hspace{-5mm}\begin{minipage}{\columnwidth}
\justifying
Continuing with the socialisation of ancillary services costs implies that smaller actors in the electricity sector would be heavily subsidising large players. The main negative consequence of this arrangement is the lack of appropriate incentives for large players to reduce their \textit{system-}
\end{minipage}

\noindent   \textit{integration cost}, that is, the cost of integrating a particular unit into the grid. Before introducing our proposed cost-causation-based framework, let us first analyse a potential argument for maintaining the status quo.

In fact, every market player benefits from a stable electricity grid: in the event of a blackout caused by a frequency excursion, consumers would not be able to purchase electricity, and generators would not be able to sell it. This could indicate that the socialisation of frequency-containment costs is the fair arrangement. However, the main problem with this setting is that some actors have no room for manoeuvre to improve the frequency security of the grid, while others do. 

To understand this issue, let us focus on under-frequency events, i.e., a frequency drop following the outage of a generator, since the unexpected disconnection of demand is a symmetrical case. There is a strong reason for making generators directly bear under-frequency services costs: if consumers bore at least a portion of these costs, while it could be argued that they do benefit from system security, they cannot act in such a way that the problem would be reduced. On the other hand, generators could be built with smaller capacities, or operate in part-loaded mode during hours when the system is under stress from a frequency perspective. This would make the frequency-containment problem less severe for all market participants, which would in turn reduce the ancillary services cost that the large generators are responsible for. Mitigation options for generators are discussed in more detail in Section~\ref{sec:Mitigation}.

Note that, eventually, demand would still cover the cost of ancillary services through updated tariffs. While grid charges for frequency services would be removed from tariffs, electricity prices would increase, since generating companies would pass on their share of the cost to consumers: by internalising their system-integration cost related to frequency security, generators would submit higher offers to the electricity market in order to recover their investment, which would translate into higher electricity prices at times. However, this does not mean that consumers would pay more for electricity. On the contrary, the goal of this cost allocation is to appropriately penalise those market actors that cause the need for ancillary services, therefore encouraging them to reduce their system-integration cost. In the end, this would lead to a reduction of total ancillary services costs, and therefore a lower charge to consumers.

Finally, it is worth pointing out that loads would be directly responsible for over-frequency ancillary services costs. The same principles discussed earlier would apply to this symmetrical case of a sudden loss of demand, which is in fact the most pressing issue in grids such as Iceland, where aluminium smelters represent the largest possible contingencies in the power system \citep{IcelandAluminium}. We touch on the topic of large consumers in Section~\ref{sec:LargeLoads}.

\subsection{How to allocate cost to different units?} \label{sec:AllocationMethod}

\begin{figure} [!t]
    \centering
    \includegraphics[width=3.5in]{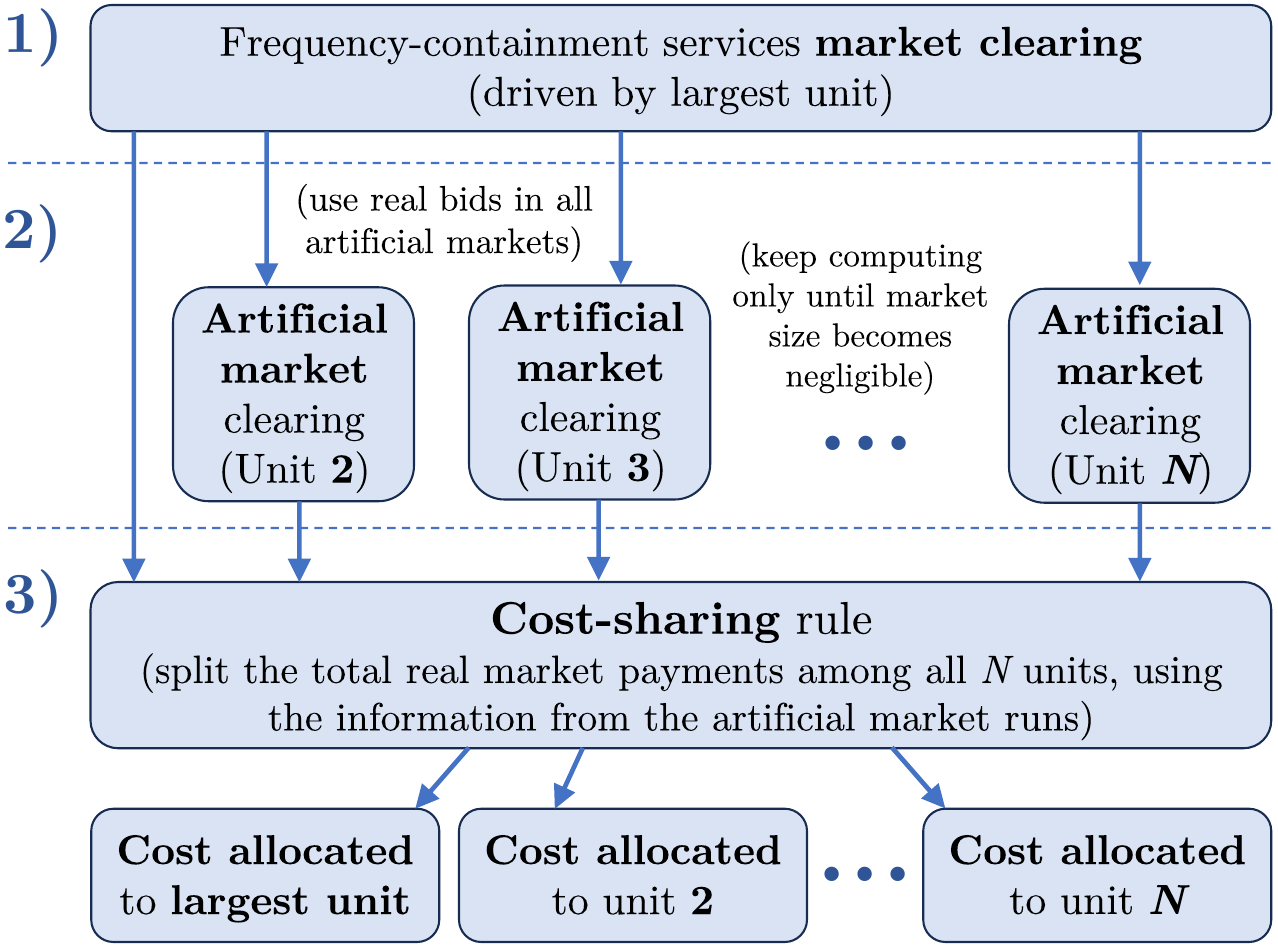}
    \caption{Three-step procedure for cost allocation.}
	\label{fig:Cost_allocation_diagrams}
\end{figure}

The allocation of ancillary services costs represents an `airport problem', i.e., it is equivalent to finding the share of investment cost in a landing strip that is attributable to each airline making use of it. While different airlines just need a runway of a certain length, the landing strip must accommodate all of them, and therefore its size is determined by the largest requirement  \citep{AirportProblem}.

We do not intend to focus on intricate mathematical formulations in the present paper. The details of the specific methodology eventually chosen to compute the cost allocation are somewhat indifferent to the discussion here, assuming that this cost allocation would indeed reflect the total cost of frequency security that each market player would create if it were the largest unit in the system. Instead, we analyse the potential consequences that such a cost allocation based on cost causation would have on future investments in the power grid. 

However, it is relevant to explain the main structure of the proposed cost allocation. The procedure, which could be applied to any given power system, is summarised in Figure~\ref{fig:Cost_allocation_diagrams}. The method is inspired by the work in \citet{CarlosAllocation}, but it is more generic to accommodate any current market structure. It includes the following steps: 
\begin{itemize}
    \item \underline{Step 1}: The ancillary services market is cleared, following the same procedure currently used by the system operator. The size of the largest unit drives the volume of ancillary services that must be procured through the market (as defined by the `$N$-1' reliability standard). As of today, this step represents the frequency reserves market, although in the future it could also include a market for inertia.
    
    \item \underline{Step 2}: A series of fictitious market simulations are run, with the same market structure as in Step 1, but now using the size of the second largest unit and any smaller unit thereafter as the drivers for the volume of services necessary. 
    The same bids as submitted to the real market are considered, since the goal is to estimate the ancillary services cost that would be created by each of the units in the system. The need for simulating these fictitious markets is demonstrated through a numerical example in Section~\ref{sec:TestCase_500MW}. 
    
    \item \underline{Step 3}: A `cost-sharing rule' is applied, with the goal of distributing the total payments to be made (determined in the real market in Step 1) among all units. The outcome of Step 2 is used for computing the cost allocated to each unit. The cost-sharing rule may be a method from cooperative game theory, left here as generic (a specific example is however given in Section~\ref{sec:CostAllocationRule}).
\end{itemize}

A numerical example to demonstrate the functioning of this cost-allocation mechanism is provided in Section~\ref{sec:TestCase}.

This mechanism essentially forces generators and loads to cover their own potential outage through the ancillary services market. The starting point is the $N$-1 criterion, meaning that the grid must be able to operate normally after a single outage of any of its $N$ units. If a generator or load wants to participate in the electricity market, they must comply with this standard, which is enforced by the system operator. A market player may not argue that it prefers forgoing the `insurance' that ancillary services provide, that is, not paying for frequency security. Since all grid assets are interconnected, a potential blackout or partial disconnection would also affect other actors in the electricity system, therefore the overall level of reliability must be decided by the regulator and system operator. 
In this sense, this is not different from the current situation, where all actors bear equally the ancillary services costs. It is simply the distribution of these costs among market players that would be updated via the cost allocation.

\subsection{Why not use a probabilistic formulation?}

One could argue that the size of the plants is not the only factor affecting frequency security: no matter how large a unit is, if it were 100\% reliable, it would not pose any frequency problem at all. In a more realistic setting, it could be argued that a large but very reliable plant would create the same challenge, at least on average, than a smaller but not as reliable unit.

In fact, \citet{GoranReserve} recommended a probabilistic approach for the cost allocation of operating reserves (note that the main difference with the present work is that a high-inertia system was implicitly assumed, where a fast frequency excursion was not a concern). To calculate the cost imputable to each generator, the authors proposed a simple method where the probability of failure of each plant is considered along with its size. However, we argue against this approach, simply because of the difficulty of its practical implementation. While in theory this technique would correctly assign blame and incentivise generators to increase their reliability, computing failure probabilities accurately is not an easy task.

Given that this probability would be used to allocate costs to specific units, this would likely be a highly contested metric, presumably leading to legal disputes between plant owners and the system operator/regulator. Failure rates are estimated using historical data on previous outages. However, these are not frequent events, only occurring at most a couple of times per year, 
therefore the probability would have to be estimated using just a few samples.
Note that only outages that occur while a unit is synchronised to the grid cause a frequency excursion, which are not very common incidents.

Not only computing current failure probabilities is challenging, but even more so auditing the improvements in reliability. The main argument in favour of the probabilistic framework would then have limited advantages, since the incentive for generators to become less prone to unexpected outages would be unclear. 

In addition, a probabilistic approach would be problematic for new plants: it would not be possible to estimate the probability of failure until a generator has been operating for some time. Investors would not have a clear view of the financial viability of their project, given that they could not estimate their cost allocation if a probabilistic formulation were used.

Furthermore, system operators currently apply the \mbox{$N$-1} criterion, meaning that frequency-containment services  
cover the worst possible (credible) contingency. Unless operators moved to a probabilistic criterion for sizing the frequency-reserves market, cost allocation would have to remain deterministic.

\subsection{When to start applying this cost allocation?} \label{sec:When}

The regulatory shift to enforce a causation-based cost allocation for ancillary services should take place as soon as possible. This new framework would shape investment in generation during the energy transition, leading the future decarbonised system to a more affordable operation, while maintaining the same levels of reliability that we enjoy today.

However, units that have already entered the market or have a regulatory agreement to do so should not be subject to this cost allocation. This is particularly relevant for nuclear plants or offshore wind farms, operational or under construction, as discussed in more detail in Section~\ref{sec:Impact}. The goal of the allocation mechanism is to 
make future plants responsible for their system-integration costs, in order to move the electricity system towards a lower need for ancillary services. However, making existing plants responsible for this cost would not contribute to achieving this goal, since there is no margin for changing the size of a generator that is already built (although certain mitigation options would be available, as explained in Section~\ref{sec:Mitigation}). The cost allocation could very negatively affect their already devised business plan, leading to likely legal actions due to the sudden change in regulation. 

In the transition period until all existing units are retired, consumers would continue subsidising existing plants for the frequency-containment costs that they create. This is still a preferable situation than for demand to subsidise all generators, and it would have an almost immediate effect in decreasing electricity tariffs.

It is worth noting that other regulatory changes have actually taken effect retroactively. For example, locational marginal pricing is currently under consideration in Great Britain \citep{REMA}, which would also affect existing assets in the power system, none of which expected this market structure when planning their business case. 
It is therefore reasonable to assume that some interim mechanism could be designed to partially penalise existing large plants, to avoid their operation at full power output during periods of low inertia.

\subsection{Why not force renewables to provide frequency support via a grid code?}

An alternative to cope with the reduction in inertia could be to force renewable generators to provide some sort of frequency support, such as synthetic inertia. It could be argued that this support has been provided for free by thermal units, therefore renewables should play their part going forward. This is in fact done through grid codes for renewable connections in some countries, particularly for wind turbines. However, this approach would lead to an overprovision of inertia, which would eventually come at a cost to consumers.

To explain this issue, let us make a comparison with investment in transmission lines: the goal is not to build a grid which is always uncongested, since that means the grid is oversized. Lack of congestion at all times implies that the investment in power lines has been higher than optimal. In a similar way, a system where every generator is forced to provide inertia would indeed be strong from a frequency-security perspective, but the investment in generation would be higher than necessary, since most renewables do not naturally provide this service. This additional cost would of course be reflected in electricity tariffs.

An economic variant of the grid code could also be devised to try to overcome this problem: instead of forcing renewables to provide inertia, they could be given the option to breach the grid code by paying a penalty. This way, rather than making the system unnecessarily strong, market forces would identify the better option between providing inertia or paying the fine. The question would then be `how to set this penalty?'

\begin{table}[t!]
  \centering
  \small
  \begin{tabular}{m{3cm} c c}
    \toprule
    & \textbf{System A} & \textbf{System B} \\
    \midrule
    Installed capacity & 2 GW & 2 GW \\
    \hspace{7.5mm}Largest unit & 1 GW & 1 GW \\
    \hspace{7.5mm}Rest of units & 1,000 $\times$ 1 MW & 2 $\times$ 500 MW \\
    \bottomrule
  \end{tabular}
  \caption{Two generic power systems.}
  \label{tab:SystemsDefinition}
\end{table}

To provide an answer, a `critical level' for inertia would have to be defined first, but this threshold would be somewhat arbitrary. When total inertia in the power grid went below the critical point, lack of it would start being penalised. The threshold could be fixed as the level of inertia at the moment of enforcing the regulation or set at a future lower level, beyond which the grid could no longer be operated safely. However, there appears no strong reason for choosing either option. 

Even if the inertia threshold were agreed upon, it would be far from clear how exactly to penalise a given renewable generator. Some heuristic would have to be used for the penalisation, as finding a rigorous procedure for cost allocation similar to the one described in Section~\ref{sec:AllocationMethod} seems implausible. On the demand side, some loads such as synchronous motors provide inertia, should other loads be penalised for not providing it? An answer to these questions would necessarily involve some degree of arbitrariness. 

Finally, we argue that, as a grid-supporting service, inertia should instead be remunerated. Inertia received no compensation in the past simply because it was not a scarce resource: if a minimum-inertia constraint were included in the market clearing, this constraint would had been inactive due to the abundance of this service, therefore its associated shadow price would have been zero. 
In the same way as constraints for transmission limits are typically included in electricity market-clearing models (leading to the concept of `congestion rent'), rigorous methods have been proposed for defining minimum-inertia constraints and computing prices from them, see for example \citet{JennyPricing}. Remunerating inertia is the approach already started by some system operators, such as in Ireland \citep{DS3Ireland} and Great Britain \citep{StabilityPathfinder}. Furthermore, renewables providing services such as synthetic inertia could indeed perceive revenues through this mechanism.

\section{Illustrative test case} \label{sec:TestCase}

Here we consider two simple power systems, System A and System B as defined in Table~\ref{tab:SystemsDefinition}, in order to illustrate the functioning of our proposed cost allocation mechanism. 

We follow the three-step methodology described in Section~\ref{sec:AllocationMethod}. Step 1 implies clearing the FCAS market (we refer to `frequency-containment ancillary services' using the same initials as in Australia, `FCAS') and therefore finding the total cost of procuring these services. Our previous work dealt with designing this first stage of the market for accommodating new services, developing methods to appropriately compensate inertia and fast frequency response as well as traditional frequency reserves \citep{JennyPricing}. Instead, the focus of the test case we include here is on the second stage of the market (Steps 2 and 3 in the methodology), that is, where should the money for these FCAS payments come from.

Regarding the model needed to simulate Steps 1 and 2, a frequency-secured Unit Commitment is typically used, i.e., a model that co-optimises energy and FCAS procurement given a certain demand and renewable generation profile. Such a model must simultaneously take into account the size of the possible largest contingency, the total inertia in the system and the speed of delivery of the frequency reserves.   
A detailed mathematical formulation for a frequency-secured Unit Commitment can be found in \citet{LuisEFR}. For the sake of clarity in the present paper, we omit these simulations and instead assume a certain numerical result for the outcome of Steps 1 and 2. This allows us to focus the discussion on the distribution of these costs among market participants.

\subsection{Cost allocation rule} \label{sec:CostAllocationRule}

\begin{figure} [!t]
    \centering
    \includegraphics[width=3.5in]{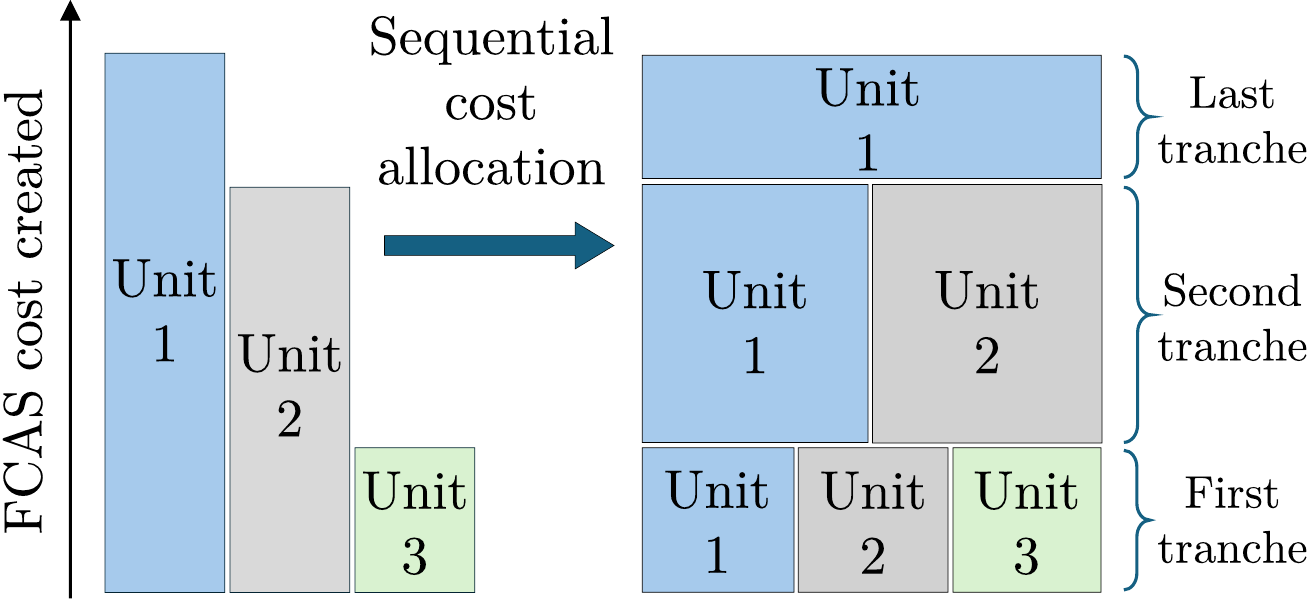}
    \caption{Sequential cost allocation in a three-unit system. All units are responsible for the first tranche of cost, all but the smallest unit for the second tranche, and just the largest unit for the last tranche.}
	\label{fig:Sequential_CostAllocation}
    \vspace{3mm}
\end{figure}

\begin{table}[t]
  \centering
  \small
  \begin{tabular}{@{}m{1.8cm} @{\hspace{2mm}}M{2cm}@{} @{\hspace{-1mm}}M{1.6cm}@{} @{}M{1.8cm}@{} @{\hspace{1mm}}M{1.6cm}@{}}
    \toprule
    \multicolumn{5}{c}{\textbf{System A}} \\
    \multicolumn{5}{c}{(Current demand: 2 GW)} \\
    \midrule
    & Power & System & FCAS cost & FCAS cost \\ 
    & produced & FCAS cost & created & allocated \\ \addlinespace[1.5mm]
    Largest unit & 1 GW & \multirow{2}{*}{\centering £10k} & £10k & £10k \\ \addlinespace[1mm]
    Rest of units & 1,000\hspace{0.4mm}$\times$\hspace{0.4mm}1\hspace{0.4mm}MW &  & £0/unit & £0/unit \\
    \midrule \midrule
    \multicolumn{5}{c}{\textbf{System B}} \\
    \multicolumn{5}{c}{(Current demand: 2 GW)} \\
    \midrule
    & Power & System & FCAS cost & FCAS cost \\ 
    & produced & FCAS cost & created & allocated \\ \addlinespace[1.5mm]
    Largest unit & 1 GW & \multirow{2}{*}{\centering £10k} & £10k & £8.4k \\ \addlinespace[1mm]
    Rest of units & 2\hspace{0.4mm}$\times$\hspace{0.4mm}500\hspace{0.4mm}MW &  & £2.5k/unit & £0.8k/unit \\
    \bottomrule
  \end{tabular}
  \caption{Cost allocation in the two generic systems.}
  \label{tab:TestCase_500MW}
\end{table}

Here we describe a specific cost-allocation rule to compute the FCAS cost that each plant is responsible for. First, let us consider the simplest rule possible, a linear cost allocation:

\vspace{-5mm}
\begin{multline}
    \textrm{FCAS cost allocated to unit } g = \\
    \hspace{10mm}= \textrm{Total system FCAS cost} \, \cdot \\
    \cdot \frac{\textrm{Power produced by unit } g}{\textrm{Sum of power produced by all generators}}
\end{multline}
    
While it is certainly preferable to enforce an FCAS cost allocation using the above proportional rule than enforcing no cost allocation at all, the analysis in \citet{CarlosAllocation} demonstrates that this still leaves room for improvement. Such a proportional rule, which only considers the power produced by each plant relative to the total power produced in the system, still maintains cross-subsidies among generators. This is due to the nonlinear impact of a large power deficit on the frequency nadir, i.e., the need for FCAS increases roughly quadratically with the contingency size (see \citet{LuisEFR} for the mathematical deduction).

Therefore, a more elaborate allocation rule is needed, which does not exhibit this undesirable property of creating cross-subsidies. For the sake of simplicity, we use here the most straightforward rule meeting this property, given by the `sequential' cost allocation \citep{littlechild1973simple}, although more intricate and potentially superior rules such as the `nucleolus' may be considered. 

In a sequential cost allocation, each unit is responsible for the additional cost that it creates to the system, as depicted in Figure~\ref{fig:Sequential_CostAllocation}. For the cases of System A and System B in Table~\ref{tab:SystemsDefinition}, the cost allocated to each of the smaller units (the 1MW units in A and 500 MW units in B) is given by:

\vspace{-5mm}
\begin{multline} \label{eq:CostAllocationRuleSmaller}
    \textrm{FCAS cost allocated to unit } g = \\
    = \frac{\textrm{FCAS cost created by unit } g}{\textrm{Total number of generators in the system}}
\end{multline}

\vspace{2mm}
\noindent And the cost allocated to the largest unit is:

\vspace{-6mm}
\begin{multline} \label{eq:CostAllocationRuleLargest}
    \textrm{FCAS cost allocated to largest unit} = \\
    \hspace{20mm}= \textrm{Total system FCAS cost} \; - \\
    - \;\textrm{FCAS cost covered by other units}
\end{multline}

\subsection{On the need to partially penalise medium-sized plants} \label{sec:TestCase_500MW}

Consider a national demand of 2 GW for both Systems A and B (we refer here to power and not energy, given that frequency is related to the instantaneous power balance). The dispatch for both systems is given in Table~\ref{tab:TestCase_500MW}, where the three last columns represent the following magnitudes:
\begin{itemize}
    \item `System FCAS cost': total cost of procuring FCAS for complying with the $N$-1 security standard. This cost is determined by the value of the largest power injection in the current system dispatch, and can be computed through a frequency-secured Unit Commitment run. It may include payments for new services such as inertia and fast frequency reserves, as well a for the traditional primary frequency reserves.
    
    \item `FCAS cost created': the cost of procuring a volume of FCAS that would cover just the contingency of a specific unit, smaller than the largest unit currently online. This cost is also obtained through a frequency-secured Unit Commitment run, but fixing the size of the contingency used within the frequency-security constraints. As many frequency-secured Unit Commitment simulations must be run as generators are in operation.
    
    \item `FCAS cost allocated': fraction of the total FCAS costs that each generator is responsible for. Obtained by applying to each unit the cost-sharing rule in either eq.~(\ref{eq:CostAllocationRuleSmaller}) or (\ref{eq:CostAllocationRuleLargest}). 
\end{itemize}

Generic values are assumed for both `System FCAS cost' and `FCAS cost created' in Table~\ref{tab:TestCase_500MW}, as explained at the beginning of this Section. Given that the largest power injection is of 1 GW for both Systems A and B, the value of the `System FCAS cost' is also the same.

The results for the `FCAS cost allocated' to each unit in System A show that the largest unit would bear all the FCAS cost, given that it is by far the largest one in the system. The much smaller 1 MW units would create a negligible FCAS, therefore no cost is allocated to them. On the other hand, in System B each of the 500 MW units does cover a portion of the overall FCAS cost, albeit significantly lower than that of the 1 GW unit. 

The reason why the 500 MW plants in System B are partially responsible for the FCAS cost is that these units are still sufficiently large to create a relevant FCAS cost if the 1 GW plant was not in operation. Even though this cost is significantly smaller than the one created by the 1 GW unit, it is not negligible. Therefore, the effect of the cost allocation mechanism is to shape the economic signal for investment in new generation: new entrants of medium size (e.g., 500 MW) would internalise their system-integration costs, that is, the potential frequency-containment costs that they would create to the overall system once the currently largest plant 
is decommissioned. On the other hand, a new unit of 1 MW would not create any FCAS cost at all, and therefore its economic signal for investment would not change. Note that if no penalisation is applied to large and medium-sized units, a new 1 GW unit (or even larger) could be built before the original one is retired, therefore maintaining the high FCAS costs for consumers.

\subsection{Incentivising flexibility in large units}

\begin{table}[t]
  \centering
  \small
  \begin{tabular}{@{}m{1.8cm} @{\hspace{2mm}}M{2cm}@{} @{\hspace{-1mm}}M{1.6cm}@{} @{}M{1.8cm}@{} @{\hspace{1mm}}M{1.6cm}@{}}
    \toprule
    \multicolumn{5}{c}{\textbf{System B}} \\
    \multicolumn{5}{c}{(Current demand: 1.5 GW)} \\
    \midrule\midrule
    \multicolumn{5}{c}{{\small First option for dispatch}} \\
    \multicolumn{5}{c}{{\small(largest unit at full output given its lowest fuel cost)}} \\
    \midrule
    & Power & System & FCAS cost & FCAS cost \\ 
    & produced & FCAS cost & created & allocated \\ \addlinespace[1.5mm]
    Largest unit & 1 GW & \multirow{2}{*}{\centering £10k} & £10k & £9.4k \\ \addlinespace[1mm]
    Rest of units & 2\hspace{0.4mm}$\times$\hspace{0.4mm}250\hspace{0.4mm}MW &  & £1k/unit & £0.3k/unit \\
    \midrule\midrule
    \multicolumn{5}{c}{{\small Second option for dispatch}} \\
    \multicolumn{5}{c}{{\small(decreasing the power output of the largest unit)}} \\
    \midrule
    & Power & System & FCAS cost & FCAS cost \\ 
    & produced & FCAS cost & created & allocated \\ \addlinespace[1.5mm]
    Largest unit & 0.9 GW & \multirow{2}{*}{\centering £7k} & £7k & £6k \\ \addlinespace[1mm]
    Rest of units & 2\hspace{0.4mm}$\times$\hspace{0.4mm}300\hspace{0.4mm}MW &  & £1.5k/unit & £0.5k/unit \\    
    \bottomrule
  \end{tabular}
  \caption{Impact of flexibility on the cost allocated to the largest unit.}
  \label{tab:TestCase_Flexibility}
\end{table}

We now focus on the incentive created by the cost allocation for flexibility in large units, i.e., for reducing power output during times of high FCAS cost (a situation typically occurring when renewable generation is high, leading to low system inertia). Consider the case in Table~\ref{tab:TestCase_Flexibility}, where we assume that the largest unit has the lowest fuel cost, and therefore it would operate at full power output if the FCAS cost allocation were disregarded.

In the `Second option for dispatch' the largest unit chooses to reduce its power output by 10\%. Given that periods of high FCAS cost are typically correlated with low electricity prices, the revenues foregone by the large unit from selling 10\% less energy would likely not be significant. However, by having this flexibility, this plant is able to significantly reduce the FCAS cost that it is allocated. At the same time, it benefits the whole system, as it reduces the $N$-1 requirement and therefore overall FCAS costs decrease. It is important to remark that the $N$-1 contingency is not driven by the rated capacity of the largest unit, but rather by the largest power injection by any unit in operation at any given time.

In summary, the benefit of the cost allocation not only lies on modifying the investment signal for large plants, but also their economic signal for operation.

\section{Impact of cost allocation on business case of generating companies and large consumers} \label{sec:Impact}

In this section, we analyse the implications for different actors in the electricity sector, including the effect on tariffs for large consumers. We identify mitigation options available to market players to reduce their allocated cost, as well as discuss the potential consequences on slowing down the progress in decarbonisation.

\subsection{Illustrative example: the British power system}

To provide some insight into the situation that investors would face if subject to this cost allocation, we use as a platform the island of Great Britain. We choose this system due to its particularly interesting characteristics: as an island with no alternating-current interconnection to other grids, it already suffers from low inertia, and the largest single contingency is of a significant size.
By the end of this decade, the $N$-1 contingency will reach up to 1.8 GW, driven by a nuclear reactor (part of a power plant with two twin reactors, currently under construction). Other significant actors from a frequency perspective include large offshore wind farms, which would drive a contingency of around 1.2 GW, due to their single-point connection to shore. 

\begin{figure} [!t]
    \centering
    \includegraphics[width=3.5in]{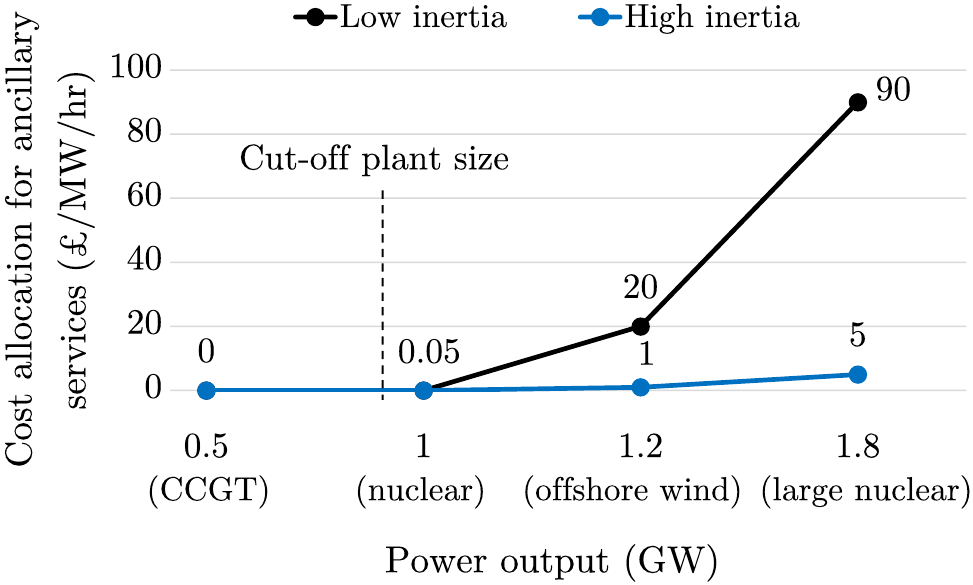}
    \caption{Illustrative example of cost allocation for each generating unit, based on the Great Britain test case in \citet{CarlosAllocation}.}
	\label{fig:Cost_allocation_curves}
\end{figure}

A cost allocation for frequency-containment services for this future British system is computed in \citet{CarlosAllocation}. From the results in said article, we obtain the curves in Figure~\ref{fig:Cost_allocation_curves}. Once the cost allocation is implemented, investors considering entering the market could conduct an equivalent study to the one in \citet{CarlosAllocation}, to understand the future cost that they would be responsible for. Figure~\ref{fig:Cost_allocation_curves} only includes two snapshots (high and low inertia hours) for a given generation mix, while the installed capacity of different technologies would evolve over time. An accurate analysis would include multi-year simulations, covering the entire lifespan of the projected plant.

A simple yet generic calculation for understanding the viability of a project is:

\vspace*{-6.5mm}
\begin{multline*}
    \sum_{y=1}^Y \;\textrm{Revenues}_{\,\textrm{electricity},\,y} \;+ \textrm{Revenues}_{\,\textrm{ancillary},\,y} \\[-6pt]
    \qquad\; –\; \textrm{Cost}_{\,\textrm{fuel},\,y} \;\;–\; \textrm{Cost}_{\,\textrm{ancillary},\,y} \;\;–\; \textrm{Cost}_{\,\textrm{others},\,y} \\[4pt]
    \quad\geq\quad \textrm{Cost}_{\,\textrm{investment}} \;\,+\;\, \textrm{Profit sought}
\end{multline*}

Where `$Y$' is the lifetime of the facility in years and `$\textrm{Cost}_\textrm{others}$' includes, e.g., maintenance costs. The differences with the current situation are:
\begin{itemize}[noitemsep]
    \item Revenues from energy could increase once generators internalise the ancillary services costs that they create, raising average electricity prices.
    \item Revenues from ancillary services could increase once services like inertia start being remunerated.
    \item Cost for ancillary services would increase for large units and decrease (or potentially disappear) for smaller ones.
    \item Fuel and investment costs could increase for large units, if the size of the plant is reduced in order to decrease its system-integration costs, therefore moving away from economies of scale.
\end{itemize}

To better understand this last point, consider a company evaluating whether to build a certain plant, with a desired capacity of 2 GW. Subject to the cost allocation, the company might consider a different design for the plant, splitting the original capacity into two different units, each of 1 GW. While the latter design may imply both a loss of efficiency (i.e., higher fuel costs for the same volume of MWh of electricity) and a higher per-MW investment, each unit would be subject to a significantly lower cost for ancillary services.

In fact, the graph in Figure~\ref{fig:Cost_allocation_curves} shows a `cut-off plant size', meaning that smaller plants would be exempt from any ancillary services payments. A cut-off point is likely to exist in all systems, given that a small volume of reserves is typically available at no cost, provided by the headroom in the marginal generators. This threshold would be highly system-specific, depending notably on the largest contingency and the usual levels of inertia, which set the required volume of frequency reserves.

Finally, for the case of the British system, the cost of frequency reserves could actually decrease in the future driven by the rise in battery storage capacity, which is the most suitable technology for providing fast frequency reserve. There is evidence that the market for this service is in fact already saturated \citep{ModoSaturation}, which is good news for large plants. Costs for other services such as inertia are however arising.

\subsection{Effect on tariffs for large consumers} \label{sec:LargeLoads}

While so far we have centred most of the discussion on large generators, large loads would be responsible for the symmetrical frequency-containment services, which include inertia and over-frequency reserves. Some examples of such loads are aluminium smelters, data centres and, in the future, electrolysers. Cross-country interconnectors may also act as big consumers, when exporting large amounts of power. 

This cost allocation would increase tariffs for large consumers, in the form of higher grid charges. 
However, some options are available specifically to loads, to reduce the frequency challenge they pose to the system. Notably, segmentation of electricity consumption within industrial facilities could be a solution: if instead of a single-point connection with the wider grid, several connections via separate circuits were in place, the facility would effectively become an aggregation of multiple loads, therefore not posing any problem in terms of large instantaneous demand losses to the overall power system. 
This modular design would of course entail a higher fixed cost. Other options available to any large actor are discussed next.

\subsection{Mitigation options for large market players} \label{sec:Mitigation}

There exist several options for large players to reduce the cost allocated to them:
\begin{itemize}
    \item For any generator: to increase their flexibility, in order to reduce power output during low inertia hours (when costs would be highest, as shown in Figure~\ref{fig:Cost_allocation_curves}).

    In particular, this cost allocation would encourage faster up and down ramping in nuclear units, and a lower minimum-stable power output. For offshore wind turbines, an already flexible technology, the cost allocation would send a market signal to decrease power output if the system is operating under low inertia.

    Currently these technologies benefit from Contracts for Difference (CfD) in Great Britain, therefore they have an incentive to produce as much electricity as possible. The cost allocation could in fact co-exist with CfD: the strike price in future CfD auctions would likely increase, since nuclear and offshore wind would not only incur higher running costs from covering a share of ancillary services, but they would also forego part of their CfD revenues by decreasing their power output at times.
    
    \item For converter-interfaced generation (e.g., offshore wind turbines): to provide grid services such as synthetic inertia. This would have a double positive effect: it would create a new revenue stream, as well as reduce the costs borne by these generators, given that overall frequency-containment costs would decrease in a grid with higher inertia.

    This service would require the much discussed grid-forming control for the power electronics converters, although so far there has been little deployment of this technology. This is because few system operators remunerate synthetic inertia yet, and grid-forming control requires some part-loading \citep{Matevosyan_PES1}, which implies a reduction in revenues from selling energy.  
    
    \item For any actor in the power system (including large consumers): to invest in external `grid-supporting assets', such as synchronous condensers. This would have the same benefits as synthetic inertia, i.e., both an increase in revenue and a reduction in cost for ancillary services. Synchronous condensers, which are essentially synchronous motors previously used just for reactive power support, are already perceiving revenues for inertia in Great Britain \citep{StabilityPathfinder}.
    
\end{itemize}

\subsection{Potential impact on decarbonisation efforts}

Two technologies would be most affected by this cost allocation: nuclear and offshore wind.
While nuclear is a highly controversial technology, it certainly does not contribute to carbon emissions. Offshore wind energy is on the other hand fully renewable, and its deployment could be hampered by the proposed cost allocation method. In contrast, polluting units such as gas plants could be exempt of any payments, as their relatively smaller power rating would likely only create a negligible cost for frequency-containment services.

While this may suggest that the regulatory change could in fact hurt decarbonisation efforts, small and clean generators would not be penalised: distributed renewables would certainly be exempt of any frequency-containment payments in any grid in the world. Therefore, the main consequence of applying the proposed cost allocation is that the business case of, for example, rooftop solar, could become more attractive in comparison to nuclear and offshore wind.

\vspace{-2mm}
\section{Outstanding questions}

Before enforcing the new regulation, certain aspects should be clarified. We highlight the following issues, where additional modelling is required to provide precise answers:
\begin{itemize}
    \item An accurate quantification of the impact on future investments is needed, for any country considering to implement a causation-based cost allocation. Generation expansion planning models that incorporate this cost allocation mechanism should be developed. It is particularly important to understand the consequences for critical technologies for decarbonisation, such as offshore wind.
    
    \item Related to the previous point, the specific mathematical method used for the cost-sharing rule needs further study. 
    The theoretical properties and computational tractability of different techniques from cooperative game theory should be further investigated, focusing on their performance in an ancillary services market setting. 
    The different options should be applied to specific electricity grids, as there may not be a single method that is superior in all instances. 
    
    \item Should we move to an $N$-2 or $N$-$k$ criterion? \citet{BialekBlackout} underscores the increased likelihood of unexpected common-mode failures, due to the rise in renewables and distributed resources. However, a level of security beyond $N$-1 might be prohibitively expensive. If it is nevertheless chosen, the cost allocation described in Section~\ref{sec:AllocationMethod} would have to be refined, to consider pairs or $k$-tuples of units.
    
    \item Although implementing a cost allocation mechanism for frequency-containment services would be a step in the right direction, other ancillary services such as reactive power, short-circuit current and system strength should also be considered \citep{Billimoria2020}. This is however not a trivial problem, given that these services are highly dependent on the topology of the network, and therefore assigning blame becomes a challenging task. 
\end{itemize}

\vspace*{-5mm}
\section{Conclusions and policy implications} \label{sec:Conclusions}

\hspace{-5mm}\begin{minipage}{\columnwidth}
\justifying
A cost allocation for frequency-containment services based on cost causation would penalise large units depending on the operational challenge they originate. Such framework would make generators and loads responsible for this share of their system-integration costs. Therefore, when planning future investments, owners would account for this cost, and potentially decide to build smaller plants, 
\end{minipage}

\vspace{1mm}
\noindent or to invest in system-supporting assets, such as synchronous condensers. An economic signal for flexibility in large units would also be created, encouraging them to reduce power output/demand during hours of high frequency-containment costs.

This would be an important change in electricity market policy, but one that would lead the future grid to a more cost-effective operation. It would have important implications for certain generating technologies, notably nuclear and offshore wind. Small consumers and generators would be the most benefitted by this new market structure, as they are currently subsidising large players. Explicit grid charges for frequency security would disappear from most tariffs, except for very large loads. Overall, electricity would become a more affordable commodity than if the status quo is maintained, given that large actors would be encouraged to reduce the `harm’ they cause to the grid.

\phantomsection
\addcontentsline{toc}{section}{CRediT authorship contribution statement}

\vspace{-2mm}
\section*{CRediT authorship contribution statement}

\textbf{Luis Badesa}: Conceptualization, Investigation (lead), Funding acquisition, Resources (lead), Visualization, Writing – Original draft, Writing – review \& editing. 

\textbf{Carlos Matamala}: Investigation (supporting). 

\textbf{Goran Strbac}: Resources (supporting).

\addcontentsline{toc}{section}{Acknowledgements}

\vspace{-2mm}
\section*{Acknowledgements}
\begin{sloppypar}
This work was funded by MICIU/AEI/10.13039/\allowbreak 501100011033 and ERDF/EU under grant PID2023-150401OA-C22. 
\end{sloppypar} 

C. Matamala was supported by the National Agency for Research and Development (ANID) through scholarship ANID/PROGRAMA BECAS CHILE
DOCTORADO under grant 2020-72210414.

\phantomsection
\addcontentsline{toc}{section}{References}

\bibliographystyle{elsarticle-harv}
\bibliography{Luis_PhD.bib}

\end{document}